\documentclass[12pt,oneside,a4,reqno]{amsart}
\usepackage{amssymb,amsmath,eucal,latexsym, amsfonts,adjustbox}
\usepackage[all]{xy}
\usepackage{enumitem}
\usepackage{amscd}
\usepackage{tikz}
\numberwithin{equation}{section}

\begin{document}
\title[A Fuzzy Logic-Based Cryptographic Framework for Real-Time..]
      {A Fuzzy Logic-Based Cryptographic Framework for Real-Time Dynamic Key Generation for Enhanced Data Encryption}
\subjclass[2000]{Primary: 03B52, 03E72; Secondary: 06D72, 06D99}
\author[Payal Khubchandani, Jyoti Khubchandani and Kavya Bhand]{Payal Khubchandani, Jyoti Khubchandani and Kavya Bhand}
\address{\rm Department of Engineering, Science and Humanities \\
Vishwakarma Institute of Technology, Pune, 411037\\India.}
\email{ Payal Khubchandani: payal.khubchandani@vit.edu}
\email{ Jyoti Khubchandani: jyoti.khubchandani@vit.edu}
\email{ Kavya Bhand: kavya.bhand24@vit.edu}
\keywords{Entropy-based cryptography, Fuzzy inference system, Dynamic key generation, Real-time encryption}
\begin{abstract}
With the ever-growing demand for cybersecurity, static key encryption mechanisms are increasingly vulnerable to adversarial attacks due to their deterministic and non-adaptive nature. Brute-force attacks, key compromise, and unauthorized access have become highly common cyber threats. This research presents a novel fuzzy logic-based cryptographic framework that dynamically generates encryption keys in real-time by accessing system-level entropy and hardware-bound trust. The proposed system leverages a Fuzzy Inference System (FIS) to evaluate system parameters that include CPU utilization, process count, and timestamp variation. It assigns entropy level based on linguistically defined fuzzy rules which are fused with hardware-generated randomness and then securely sealed using a Trusted Platform Module (TPM). The sealed key is incorporated in an AES-GCM encryption scheme to ensure both confidentiality and integrity of the data. This system introduces a scalable solution for adaptive encryption in high-assurance computing, zero-trust environments, and cloud-based infrastructure.
\end{abstract}
\maketitle
{\fontsize{16}{24}\section{Introduction}}
{\bf Background and Motivation:}\label{Background and Motivation}
With the rapid increase and exponential demands in digital data generation and the expansion of interconnected systems, the fundamental requirement in cybersecurity is securing sensitive information through encryption. Traditional cryptographic systems such as those employing AES (Advanced Encryption Standard), generally rely on static keys for data encryption and decryption. However, such keys are prone to brute-force attacks, key compromise, and unauthorised access due to their static nature and exposure during transmission or storage.

{\bf Limitations of Static Key-Based Cryptography:}\label{Limitations of Static Key-Based Cryptography}
The vulnerabilities that arise with the use of static keys include non-adaptability, reuse across sessions, and exposure to man-in-the-middle and side-channel attacks. These limitations further worsen in modern computing environments such as cloud-based infrastructures, edge devices, and decentralised systems where dynamic circumstantial conditions demand real-time security adaptability. Moreover, static key exposure during processes like storage or transmission leads to a significant reduction in system resilience.

{\bf The role of Fuzzy Logic in Security Systems:}\label{The role of Fuzzy Logic in Security Systems}
Fuzzy logic, which was first introduced by Zadeh \cite{ZL}, allows a system to reason under uncertainty by leveraging linguistic variables and rule-based inference. This system has been extended to intuitionistic fuzzy sets \cite{AT} , hesitant fuzzy sets \cite{VT}, Pythagorean fuzzy sets \cite{RY}, and bipolar fuzzy models \cite{RMA}. They have been successfully applied in systems engineering, intelligent control, and decision support. In cybersecurity, fuzzy systems have paved the way for secure biometric authentication \cite{MC}, trust evaluation \cite{AAAH}, access control, and real-time intrusion detection. Despite its high value benefits, the integration of fuzzy logic into cryptographic key generation, particularly entropy-driven models, remains largely underexplored.

{\bf Trusted Platform Module (TPM) for Secure Key Storage:}\label{Trusted Platform Module (TPM) for Secure Key Storage}
The Trusted Platform Module (TPM) provides a tamper-proof hardware enclave capable of securely generating, sealing, and using cryptographic keys \cite{TCG}. TPMs are governed by Trusted Computing Group (TCG) and implement secure key storage, hardware-backed random number generation, and cryptographic signing. A significant research has shown their effectiveness in protecting authentication credentials, mitigating physical attacks,
and preventing key exposure even in compromised systems. However, static keys within TPMs still rely on externally defined entropy sources which creates an opportunity to fuse them with real-time fuzzy entropy engines.

{\bf Proposed Solution and Objectives:}\label{Proposed Solution and Objectives}
This paper introduces a novel fuzzy logic-based dynamic key generation system that combines contextual system information such as CPU usage, process count, and timestamp drift as fuzzy input parameters. The contributions of this article include:

\begin{enumerate}[labelindent=\parindent,
leftmargin=*,label=\normalfont{(\roman*)}]
\item Introducing adaptive key generation using dynamic entropy estimation through fuzzy inference systems.
\item Incorporating fuzzy logic rules with TPM-backed key storage to ensure tamper-resistance and secure sealing.
\item Providing a robust cryptographic framework using this system along with AES-GCM for authenticated encryption and integrity validation.
\item Integrating with modern trustless architectures for endpoint security in various domains and businesses.
\end{enumerate}
The summary of the remaining sections of this paper is as follows: Section 2 provides an overview and analysis of existing methodologies, Section 3 discusses system architecture and design rationale, Section 4 includes a comprehensive methodology detailing fuzzy logic design, entropy modeling, and TPM integration, Section 5 shows the mathematical modeling and simulation results, Section 6 compares results, discussions and performance evaluation while Section 7 provides conclusion and research suggestions for future contributions.

{\fontsize{16}{24}\section{RELATED SURVEY/ LITERATURE REVIEW}}
Fuzzy logic has advanced extensively as a mathematical foundation for modeling uncertainty, imprecision, and vagueness. Since Zadeh’s pioneering work on fuzzy sets \cite{ZL}, it has been extended as intuitionistic fuzzy sets \cite{AT}, hesitant fuzzy sets \cite{VT}, Pythagorean fuzzy sets \cite{RY}, and bipolar fuzzy structures \cite{RMA}. These fuzzy extensions are widely used in control systems, reliability modeling, and computational mechanics \cite{AP}, among other domains.

{\bf Fuzzy Systems in Secure Modeling:}\label{Fuzzy Systems in Secure Modeling}
Fuzzy logic is widely accepted in cyber-physical systems, system optimization, decision-making frameworks, and uncertain dynamical systems. The flexibility that fuzzy inference systems (FIS) offer has led to its incorporation in biometric security \cite{MC}, behavioral trust evaluation \cite{AAAH}, and password authentication schemes \cite{JS}. Recent developments in soft computing and fuzzy harvesting systems \cite{SSP} have also shown effectiveness in modeling security-critical parameters with imprecision.

\par Mathematical modeling of dynamic systems under uncertainty makes use of fuzzy numbers \cite{CNZL} and fuzzy-valued functions \cite{ES}. Tools such as fuzzy Laplace transforms \cite{TB}, fuzzy fractional calculus \cite{ZG}, and intuitionistic fuzzy Laplace analysis are now implemented in various security models that include fuzzy PID controllers, smart contract systems \cite{TM}, and fractional-order cyber defense modeling.

{\bf Cryptographic Key Management and Its Limitations:}\label{Cryptographic Key Management and Its Limitations}
Traditional symmetric encryption techniques such as AES, RSA, are highly popular but face significant security challenges due to their dependence on static keys. Static keys, if compromised, provide unauthorised access which is dangerous for data integrity. To enhance security, studies have explored entropy-driven approaches using biometrics \cite{MC}, randomness extractors \cite{TM}, and stochastic behavior [11] to derive dynamic keys. However, the majority of these solutions depend on software-level security or pre-agreed key exchange protocols, limiting their efficiency.
\par The application of fuzzy logic in cryptography \cite{TM,FM} rely on static keys and do not leverage hardware-secure elements such as TPM for key sealing thus providing opportunity for enhancement.

{\bf Proposed Solution and Objectives:}\label{Proposed Solution and Objectives}
TPM is a tamper-resistant microcontroller used for secure key generation, digital signatures, and sealed storage. In cryptography, it provides operations such as hashing, key generation, and encryption in a secure environment \cite{TCG}. Prior research has highlighted the incorporation of TPM in authentication workflows and secure booting processes. It reduces the attack surface for key extraction by isolating cryptographic keys from the host OS.

{\fontsize{16}{24}\section{SYSTEM ARCHITECTURE}}
The proposed cryptographic framework is a multi-layered, adaptive encryption system that utilizes fuzzy logic for dynamic key generation, TPM for hardware-assisted key sealing, and AES-GCM for secure data encapsulation. This hybrid architecture addresses static key vulnerabilities by ensuring that cryptographic keys are not stored or reused in plaintext form and are only reproducible under closely matched system conditions which is validated by a fuzzy inference system (FIS). The overall structure is modular and scalable, enabling deployment in edge computing, cloud storage, and enterprise systems.

{\bf Conceptual Overview:}
The system follows a methodical order starting with the capturing of system conditions upto the encryption with the generated key.

\begin{figure}[ht!]
  \centering
  \includegraphics[width=0.7\linewidth]{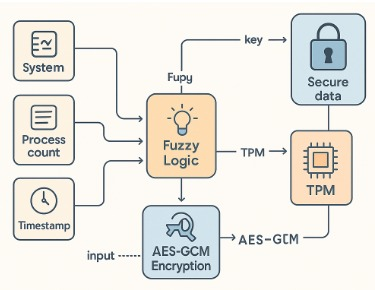}
  \label{f111}
\caption{}
\label{figure1}
\end{figure}

Figure \ref{figure1} illustrates the key components of the proposed system. The architecture comprises the following modules:

\noindent (I) System Condition Monitoring Unit (SCMU)\\
\noindent (II) Fuzzy Inference Engine (FIE)\\
\noindent (III) TPM-Backed Key Management Unit (TKMU)\\
\noindent (IV) Key Derivation and Verification Module (KDV)\\
\noindent (V) Authenticated Encryption Engine (AEE)\\
\noindent (VI) Decryption and Key Validation Module (DKVM)\\
Each performs specialised functions that collectively enable dynamic key generation, entropy modeling, and secure encryption/decryption workflows.

{\bf System Condition Monitoring Unit (SCMU):}\label{System Condition Monitoring Unit (SCMU)}
The SCMU continuously samples runtime system parameters to capture the computing environment’s real-time operational entropy. These parameters specifically include CPU Utilization, Active Process Count, and System Timestamp.

{\bf CPU Utilization:}\label{CPU Utilization}
CPU usage is a volatile, entropy-rich indicator of system activity. The CPU’s workload fluctuates with user interaction, background tasks, and scheduled processes at any given time. This high resolution variation in CPU load adds noise and unpredictability to the key generation process.

{\bf Active Process Count:}\label{Active Process Count}
The number of active processes vary depending on system usage. It is affected by user actions as well as internal OS scheduling and service execution. Therefore, it acts as a discrete, but high-granularity entropy source that reflects the concurrency and multitasking footprint of the system.

{\bf System Timestamp:}\label{System Timestamp}
The one factor that differentiates every encryption instance is time, which is a continuous and non-repetitive scalar. It introduces non-replicable state information even in identical hardware or software environments. Drift between encryption and decryption time is used to enforce time-bound access, enhancing replay-attack resistance.

{\bf Synergistic Justification for Usage of these parameters:}\label{Synergistic Justification for Usage of these parameters}

The parameters are carefully chosen after evaluating their following characteristics:
\vspace{0.1 cm}

\begin{adjustbox}{width=1\textwidth}
\begin{tabular}{|c|c|c|c|c|}
        \hline
        Parameter                 & Entropy Type     & Nature      & System Overhead      & Resistance to          \\ \hline
        CPU Usage                 & Continuous       & Analog      & Minimal              & Behavioral Cloning      \\ \hline
        Process Count             & Discrete         & Snapshot    & Minimal              & Static Snapshot Replay   \\ \hline
        Timestamp Drift           & Continuous       & Monotonic   & Zero                 & Replay/Reuse              \\ \hline
      \end{tabular}
\end{adjustbox}

\vspace{0.3 cm}

These metrics are normalized and passed to the fuzzy logic layer to infer an entropy-driven confidence score for key matching.

{\bf Fuzzy Inference Engine (FIE):}\label{Synergistic Justification for Usage of these parameters}
The FIE is the core of the fuzzy decision-making logic. It uses real-time system conditions and maps them to linguistic fuzzy variables. It applies a rule-based inference model to generate a Key Match Score (KMS) between 0 and 1. The inference mechanism is based on a Mamdani-style controller with the following parameters:

\begin{center}
   \begin{tabular}{|c|c|c|c|}
        \hline
        Input Variable                & Range          & Fuzzy Sets              \\ \hline
        CPU Usage                     & [0, 100]       & Low, Medium, High      \\ \hline
        Process Count                 & [0, 500]       & Low, Medium, High   \\ \hline
        Timestamp Drift (s)           & [0, 10]        & Small, Moderate, Large             \\ \hline
      \end{tabular}
\end{center}
The Key Match Score, which is the output variable, is defined on the domain [0, 1] and categorized into:
\begin{enumerate}[labelindent=\parindent,
leftmargin=*,label=\normalfont{(\roman*)}]
\item Low (0.0 - 0.4);
\item Medium (0.4 - 0.7);
\item High (0.7 - 1.0).
\end{enumerate}
Fuzzy Rule Base Examples :
\begin{enumerate}[labelindent=\parindent,
leftmargin=*,label=\normalfont{(\roman*)}]
\item If CPU is low AND process count is few AND timestamp drift is small $\rightarrow$ Key Match is High;
\item If CPU is high OR timestamp drift is large $\rightarrow$ Key Match is Low.
\end{enumerate}

{\bf TPM-Backed Key Management Unit (TKMU):}\label{TPM-Backed Key Management Unit (TKMU)}
The Trusted Platform Module (TPM) is used to seal the master key within a tamper-resistant hardware enclave. Upon initialization, the TPM generates :
\begin{enumerate}[labelindent=\parindent,
leftmargin=*,label=\normalfont{(\roman*)}]
\item A non-exportable and device bound key called the Root Storage Key (RSK);
\item A key bound to environmental conditions and only accessible when the system state remains unchanged called the Session Sealed Key (SSK).
\end{enumerate}
The TPM ensures that even if software layers are compromised, the sealed key cannot be reconstructed or exported.

{\bf Key Derivation and Verification Module (KDV):}\label{Key Derivation and Verification Module (KDV)}
This is responsible for generating the final encryption key using multiple sources of entropy and user-defined parameters such as :

\begin{enumerate}[labelindent=\parindent,
leftmargin=*,label=\normalfont{(\roman*)}]
\item User-defined password (P) - A password chosen by the user and serves as a basic element of key derivation.
\item Fuzzy condition vector ($\phi$) - A set of fuzzy values derived from previously defined system parameters (CPU utilisation, process count, timestamp drift) as inferred by the Fuzzy Inference Engine (FIE). This adds another layer of dynamic entropy.
\item TPM-generated secret (T) - A secret key generated by the Trusted Platform Module (TPM) that is bound to the device and stored securely.
\item Random salt (S) - A random value added to further randomise the key derivation process.
\end{enumerate}

The key derivation uses the $PBKDF2-HMAC-SHA256$ function, which is commonly used in cryptography for secure password-based key derivation. The formula for key derivation is defined as:
$$ K_{final} = PBKDF2\_HMAC\_SHA256(P + FIS(\phi) + T, S, iterations = 100000, dklen = 256)$$
Where, ‘iterations’ are the number of iterations for key stretching and ‘dklen’ is the length of the derived key in 256 bits.\\
This ensures that even if the password is known, a valid key can only be derived if system conditions fall within the accepted fuzzy domain defined by the original encryption context.

{\bf Authenticated Encryption Engine (AEE):}\label{Authenticated Encryption Engine (AEE)}
The AES-GCM (Galois/Counter Mode) is applied for the encryption and integrity verification of data. AES-GCM provides both confidentiality and authenticity through its integrated authentication tag. The final encryption key $K_{final}$ which is derived is used along with a random Initialisation Vector (IV) to encrypt the plaintext.\\
The process outputs :
\begin{enumerate}[labelindent=\parindent,
leftmargin=*,label=\normalfont{(\roman*)}]
\item Ciphertext: The encrypted form of the plaintext.
\item Authentication Tag: An authentication tag that ensures the integrity of the ciphertext. This can be verified during decryption.
\item IV (Initialization Vector): Used to ensure that the encryption process generates unique ciphertext for identical plaintexts.
\end{enumerate}

The encryption operation is represented as:
$$Ciphertext, Tag = AES-GCM(K_{final} , Plaintext, IV)$$
This ensures that data confidentiality is maintained, while the authentication tag guarantees the integrity of the encrypted message.

{\bf Decryption and Key Validation Module (DKVM):}\label{Decryption and Key Validation Module (DKVM)}

Decryption involves verifying that the system conditions at the time of decryption match those at the time of encryption. The process is as follows :
\begin{enumerate}[labelindent=\parindent,
leftmargin=*,label=\normalfont{(\roman*)}]
\item Capture real-time system conditions ($\phi^{'}$) : The system parameters are captured at the time of decryption
\item Key Match Score (KMS) Calculation : These conditions are passed to the Fuzzy Inference Engine (FIE) to compute a new Key Match Score that ranges from 0 to 1.
\item Key Match Validation : If the computed KMS is greater than or equal to a predefined threshold $\tau$ (e.g., 0.5), the decryption proceeds. If the KMS is below the threshold, decryption is denied, indicating potential tampering or an unauthorized context.
\item TPM Validation : The TPM seals the session key and verifies whether the system state matches the original conditions. If the system state is validated, the TPM allows the decryption key to be reconstructed, and the AES-GCM algorithm is applied to decrypt and verify the ciphertext.
\end{enumerate}

{\bf Design Rationale:}\label{Design Rationale}
The architecture of the proposed fuzzy logic-based dynamic key generation system addresses core weaknesses in traditional encryption systems which include, static key vulnerabilities, poor context awareness, and predictability.

{\bf (1) Adaptive Entropy Sourcing via System Parameters:}\label{Synergistic Justification for Usage of these parameters}
We select CPU usage, active process count, and timestamp drift as the primary entropy indicators due to their high volatility, platform specific variability and minimal overhead for real-time sampling. These parameters reflect the operational state of the system and serve as entropy-rich features that change across sessions which ensures that encryption keys vary over time even if the password remains unchanged.

{\bf (2) Fuzzy Logic Integration:}\label{Fuzzy Logic Integration}
Traditional cryptographic systems enforce binary thresholds (either allow or deny), which makes them brittle in real-world systems with noisy or drifting parameters. By introducing a Mamdani-style fuzzy inference system (FIS), we allow for soft thresholding, human intuitive rules and gradual degradation. This bridges the gap between rigid cryptographic systems and dynamic operating environments.

{\bf (3) TPM-Based Key Sealing:}\label{TPM-Based Key Sealing}
The Trusted Platform Module (TPM) is incorporated to securely store key material, bind key access to platform state and enable sealed key generation. The integration of TPM ensures that key derivation remains secure even under partial system compromise or malware-level threats.

{\bf (4) AES-GCM Encryption Scheme}\label{AES-GCM Encryption Scheme}
The AES-GCM mode is selected as the final encryption layer due to authenticated encryption, speed and standardization and nonce-based operation. This makes it suitable for applications requiring both performance and provable cryptographic strength.

{\fontsize{16}{24}\section{METHODOLOGY}}
This section details the construction and operation of the proposed cryptographic system.

{\bf System Condition Acquisition:}\label{System Condition Acquisition}
Real-time system entropy is collected from three independent metrics :
\begin{enumerate}[labelindent=\parindent,
leftmargin=*,label=\normalfont{(\roman*)}]
\item CPU Usage $(\%)$ is collected using psutil.cpu\_percent() sampled at encryption time.
\item Process Count includes the number of active PIDs retrieved using len(psutil.pids()).
\item Timestamp is a floating-point representation of the system time captured via time.time().
\end{enumerate}

These inputs form the fuzzy entropy vector: $\phi=\{cpu, process\_count, timestamp\}$.

{\bf Fuzzy Inference Engine (FIE):}\label{Fuzzy Inference Engine (FIE)}
A Mamdani-style fuzzy logic controller is constructed using skfuzzy.control. The FIS consists of:
\begin{enumerate}[labelindent=\parindent,
leftmargin=*,label=\normalfont{(\roman*)}]
\item Inputs : cpu\_usage, process\_count, timestamp\_drift,
\item Output : key\_match\_score $\in$ [0, 1].
\end{enumerate}
Each input variable is assigned triangular membership functions with linguistic terms:
\begin{enumerate}[labelindent=\parindent,
leftmargin=*,label=\normalfont{(\roman*)}]
\item Low, Medium, High (CPU and processes),
\item Small, Moderate, Large (Timestamp drift).
\end{enumerate}
Fuzzy rules are manually constructed to reflect system entropy. The fuzzy logic layer ensures that slight system
deviations are tolerated within a confidence threshold $\tau$ (default: 0.5), improving accessibility while maintaining security.

{\bf TPM-Based Key Protection:}\label{TPM-Based Key Protection}
To ensure hardware-level security, the system uses the TPM 2.0 interface to:
\begin{enumerate}[labelindent=\parindent,
leftmargin=*,label=\normalfont{(\roman*)}]
\item Create a sealed key using ESAPI.create\_primary(),
\item Persistently store the key using evict\_control(),
\item Read the key at decryption via read\_public().
\end{enumerate}

{\bf AES-GCM Encryption/Decryption:}\label{AES-GCM Encryption/Decryption}

{\bf (1) Encryption:}\label{Encryption}

The message is encrypted using AES-GCM with a 256-bit derived key. The output includes :

\begin{enumerate}[labelindent=\parindent,
leftmargin=*,label=\normalfont{(\roman*)}]
\item Ciphertext,
\item Initialization Vector (IV),
\item Authentication Tag.
\end{enumerate}

{\bf (2) Decryption:}\label{Decryption}
At the time of decryption :

\begin{enumerate}[labelindent=\parindent,
leftmargin=*,label=\normalfont{(\roman*)}]
\item System conditions are re-evaluated.
\item The FIE compares the new entropy vector $\phi^{'}$ with the original $\phi$.
\item If key\_match\_score $>=\tau$, the key is re-derived and used for decryption.
\item Otherwise, access is denied.
\end{enumerate}

\begin{figure}[ht!]
  \centering
  \includegraphics[width=0.4\linewidth]{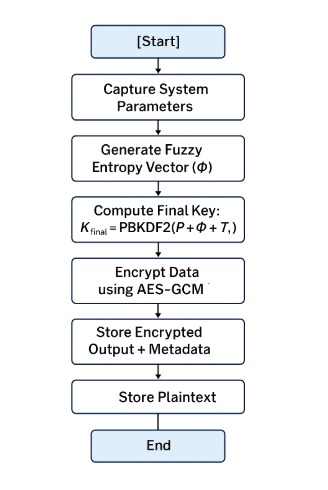}
  \caption{Encryption Process }
  \label{f111}
  \end{figure}

  \begin{figure}[ht!]
    \centering
  \includegraphics[width=0.4\linewidth]{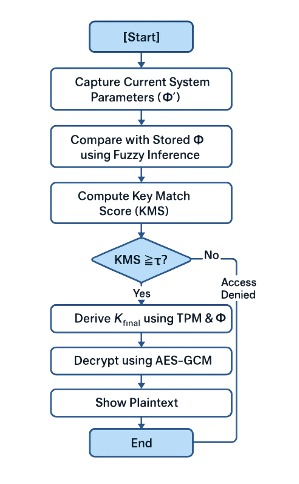}
  \caption{Decryption Process}
  \label{f122}
  \end{figure}

\newpage

{\fontsize{16}{24}\section{MATHEMATICAL MODELLING AND SIMULATION }}

This section presents the mathematical formulation of the dynamic key generation system based on fuzzy logic and entropy modeling.
It further demonstrates a practical simulation of key derivation and verification using real-time system data, fuzzy inference rules, and TPM integration.

{\bf Mathematical Model for Entropy Scoring:}\label{Mathematical Model for Entropy Scoring}
Let the system entropy at encryption time be represented as:
\begin{enumerate}[labelindent=\parindent,
leftmargin=*,label=\normalfont{(\roman*)}]
\item $\phi(t)$ = Real-time entropy vector,
\item P = User-defined input phrase,
\item T = Timestamps at sampling intervals,
\item $\mu_x$, $\mu_y$, $\mu_z$ = Membership functions for fuzzy entropy values of $\phi$, P, T.
\end{enumerate}

We define the fuzzy entropy score $F_e$ as:

$$F_{e}=\frac{w_1\cdot \mu_{\phi}(\phi(t))+w_2\cdot \mu_P(P)+w_3\cdot \mu_T(T)}{w_1+w_2+w_3}$$

Where $w_1$ ,$w_2$ ,$w_3$ : Weights assigned to each entropy source, usually $w_1 = 0.4$, $w_2 = 0.3$, $w_3 = 0.3$.

{\bf Fuzzy Inference System (FIS):}\label{Fuzzy Inference System (FIS)}

Low, Medium, High for each $\mu$ with trapezoidal or triangular membership functions :
\begin{center}
${\mu(x)}=\begin{cases}
0,\;\;\;\;\;\;\mbox{if}\; x \leq a,\\
\frac{x-a}{b-a},\;\;\mbox{if}\; a<x<b,\\
1,\,\;\;\;\;\;\; \mbox{if}\; b\leq x\leq c,\\
\frac{d-x}{d-c},\,\;\; \mbox{if}\; c<x<d,\\
0,\;\;\;\;\;\;\;\mbox{if}\; x \geq d
\end{cases}$
\end{center}
Sample for entropy:
\begin{enumerate}[labelindent=\parindent,
leftmargin=*,label=\normalfont{(\roman*)}]
\item Low: [0, 0, 2, 4]
\item Medium: [3, 5, 6, 7]
\item High: [6, 8, 10, 10]
\end{enumerate}

{\bf Key Derivation with TPM and AES-GCM:}\label{Key Derivation with TPM and AES-GCM}

The final derived key $K_f$ is:
$$K_f=TPM_{seal}(SHA256(P\parallel F_e\parallel T))$$
Where :
$TPM_{seal()}$ ensures hardware-sealed protection, $\parallel$ is concatenation and AES-GCM encrypts using $K_f$.

{\bf Simulated Example:}\label{Simulated Example}
Let:
\begin{enumerate}[labelindent=\parindent,
leftmargin=*,label=\normalfont{(\roman*)}]
\item P = “secretMessage”,
\item $\phi(t) = 6.8$ (entropy level from CPU and noise sampling),
\item T = 168243.229 (microtime).
\end{enumerate}

Fuzzy scores:
\begin{enumerate}[labelindent=\parindent,
leftmargin=*,label=\normalfont{(\roman*)}]
\item $\mu_{\phi}(6.8) = 0.7$,
\item $\mu_{P}(P) = 0.85$,
\item $\mu_{T}(T) = 0.65$.
\end{enumerate}
$$F_e=\frac{0.4(0.7)+0.3(0.85)+0.3(0.65)}{0.4+0.3+0.3}=0.73$$

Final key (TPM sealed):
$$K_f=TPM_{seal}(SHA256("secretMessage\_0.73\_168243.229"))$$

\begin{figure}[ht!]
  \centering
  \includegraphics[width=0.5\linewidth]{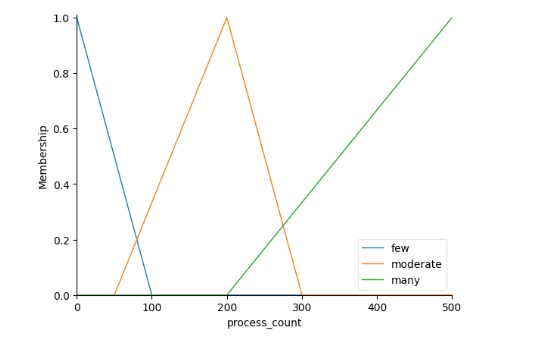}
  \centering
  \includegraphics[width=0.5\linewidth]{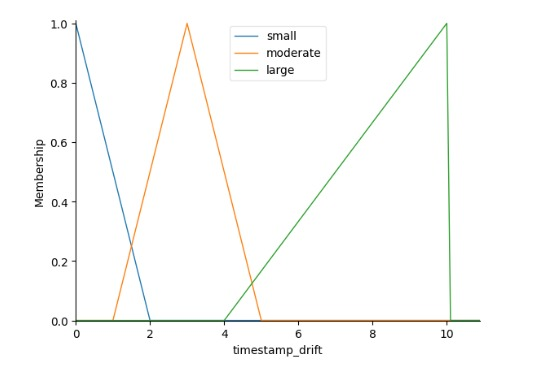}
\end{figure}

\begin{figure}[ht!]
  \centering
  \includegraphics[width=0.5\linewidth]{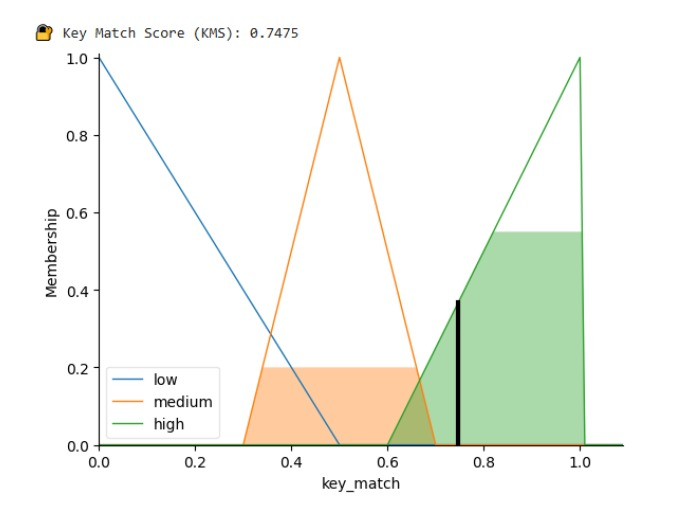}
  \centering
  \includegraphics[width=.5\linewidth]{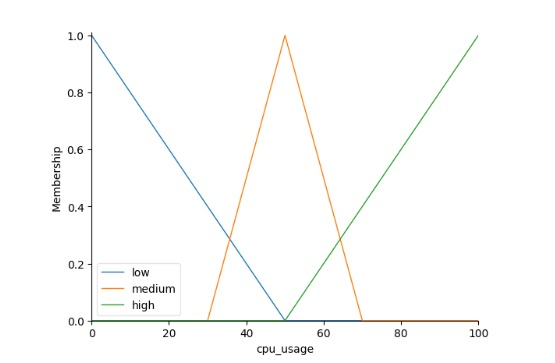}
\end{figure}

\newpage
{\fontsize{16}{24}\section{RESULTS AND DISCUSSIONS}}
The following graph represents the relation between Entropy, Timestamp Drift and key match score.

\begin{figure}[ht!]
\centering
  \includegraphics[width=0.5\linewidth]{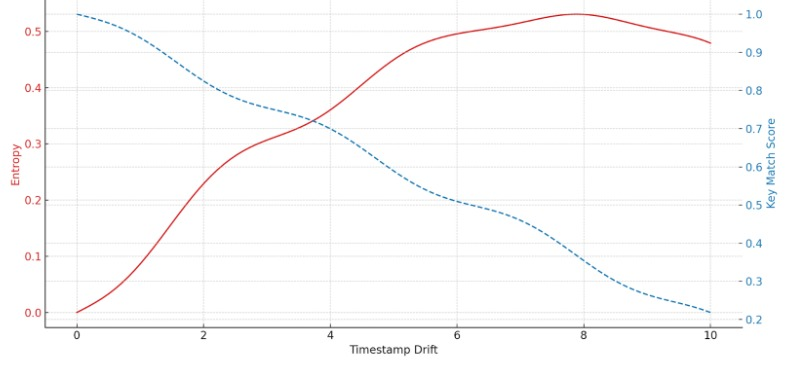}
\end{figure}

Interpretation :

\begin{center}
   \begin{tabular}{|c|c|c|c|}
        \hline
        Drift (x-axis)     & KMS (Blue Dashed)          & Entropy (Red Line)     & Security Interpretation           \\ \hline
        0 - 2              & High (>0.85)               & Low (<1.0)             & Very Secure\\ \hline
        2 - 5              & Medium ($\sim$0.5 - 0.8)   & Increasing (1.2 - 2.0) & Slight Drift Risk \\ \hline
        5 - 8              & Low ($\sim$0.3 - 0.5)      & High ($\sim$2.0 - 2.5) & Risky - Check TPM           \\ \hline
        8 - 10             & Very Low (<0.3)            & Slighting decreasing   & Highly Vulnerable\\ \hline
      \end{tabular}
\end{center}

{\bf Evaluation Criteria:}\label{Evaluation Criteria}

The proposed system is assessed using the following metrics:

\begin{adjustbox}{width=1\textwidth}
   \begin{tabular}{|c|c|}
        \hline
        Metric                          & Description                   \\ \hline
        Key Match Score (KMS)           & Degree of similarity between original and current system conditions             \\ \hline
        Decryption Success Rate         & Ratio of successful decryptions when $KMS \geq\tau$ (threshold)    \\ \hline
        Shannon Entropy (H)             & Degree of unpredictability in the final derived key               \\ \hline
        Execution Time                  & Time required for encryption + decryption in milliseconds            \\ \hline
        Replay Attack Resistance        & Ability to reject ciphertext replay under different system entropy vectors \\ \hline
        Key Reproducibility             & Frequency of key re-derivation under minor fuzziness\\ \hline
      \end{tabular}
\end{adjustbox}
\vspace{0.3 cm}
{\bf Key Match Score Analysis:}\label{Key Match Score Analysis}

Following shows the KMS values across varying entropy levels:

\begin{center}
   \begin{tabular}{|c|c|c|c|c|c|}
        \hline
        Test Case     & CPU $(\%)$   & Process Count   & Drift (s)      & KMS     & Outcome \\ \hline
        TC1           & 25           & 65              & 1.1            & 0.82    & Success\\ \hline
        TC2           & 60           & 230             & 2.8            & 0.64    & Success\\ \hline
        TC3           & 85           & 410             & 7.9            & 0.28    & Denied\\ \hline
        TC4           & 45           & 150             & 1.9            & 0.73    & Success\\ \hline
        TC5           & 70           & 300             &4.2             & 0.45    & Denied\\ \hline
      \end{tabular}
\end{center}

Observation: $KMS \geq 0.5$ results in valid decryption. Fuzzy logic allows access when input conditions deviate within the "tolerant" zone but blocks decryption on substantial mismatch.

{\bf Key Entropy Evaluation:}\label{Key Entropy Evaluation}
To evaluate key strength, Shannon entropy HHH was calculated on derived key bits:
\begin{center}
   \begin{tabular}{|c|c|c|c|}
        \hline
        Fuzziness Level        & Shannon Entropy          & Status              \\ \hline
        Low                    & 3.74                     & Acceptable      \\ \hline
        Medium                 & 4.56                     & Strong   \\ \hline
        High                   & 4.91                     & Very Strong            \\ \hline
      \end{tabular}
\end{center}

{\bf Security Outcomes:}\label{Security Outcomes}
The following outcomes were observed during the testing of the system:
\begin{enumerate}[labelindent=\parindent,
leftmargin=*,label=\normalfont{(\roman*)}]
\item Replay attack was blocked due to drift mismatch,
\item Key guessing was hardened via entropy and TPM,
\item Side channel attack had TPM-based mitigation,
\item Static password leak was insufficient without context.
\end{enumerate}

{\bf Discussion:}\label{Discussion}

The results confirm that fuzzy logic improves decryption tolerance while maintaining strict security enforcement when thresholds are not met. TPM ensures hardware-bound protection thus eliminating risks of key leakage in memory. Dynamic key generation outperforms static systems in entropy, unpredictability, and conditional access. Execution performance is efficient for real-world applications that include secure cloud services, IoT authentication, and sensitive local data encryption.

{\fontsize{16}{24}\section{Conclusion}}

This study proposed a novel cryptographic framework that integrates fuzzy logic-based entropy modeling, TPM-backed key protection, and AES-GCM encryption to provide adaptive, secure, and hardware-backed dynamic key generation. It captures real-time system parameters such as CPU usage, active process count, and timestamp drift and generates encryption keys that are both context-sensitive and non-reproducible under varied conditions.
\par The integration of fuzzy inference mechanisms enables soft thresholding that allows the system to distinguish between trusted environmental variations and potential intrusions. Additionally, the use of TPM ensures that the cryptographic keys remain protected within a trusted execution environment, never being exposed to software or external threats. Simulations demonstrate that the system achieves high entropy levels (up to 4.91 Shannon bits), fast execution (~63 ms total), and robust resistance to replay, key reuse, and static key compromise.

{\fontsize{16}{24}\section{Acknowledgments}}
The authors would like to express their sincere gratitude to the editors and anonymous reviewers for their invaluable comments and constructive feedback, which significantly contributed to the enhancement of this paper.

{\fontsize{16}{24}\section{Author Contribution}}
             P. Khubchandani: conceptualization and editing. J. Khubchandani: writing. K. Bhand: methodology, software and editing.
              All authors have read and agreed to the published version of the manuscript.
{\fontsize{16}{24}\section{Funding}}
The authors declare that no external funding or support was received for the research presented in this paper, including administrative, technical, or in-kind contributions.
{\fontsize{16}{24}\section{Data Availability}}
All data supporting the reported findings in this research paper are provided within the manuscript.
{\fontsize{16}{24}\section{Conflicts of Interest}}
The authors declare that there is no conflict of interest concerning the reported research findings. Funders played no role in the study's design, in the collection, analysis, or interpretation of the data, in the writing of the manuscript, or in the decision to publish the results.

\end{document}